# A Model for Crosstalk in Micropattern Gas Detectors

Marcus Hohlmann

*Abstract*–Crosstalk characteristics such as pulse amplitude and shape are studied with a simple PSpice model of the capacitive couplings within an MPGD. The crosstalk pulse shape can be understood as due to a CR-differentiator. Crosstalk can occur simultaneously through more than one capacitive coupling path. The crosstalk signals on these paths add differently if the signal is induced via a current source as in normal detector operation or via a voltage source as is often done in benchtop tests with an external voltage pulse generator. A few means for reducing the crosstalk are investigated with the model. It shows that a low-impedance AC path from the amplification electrode that faces the readout structure to ground is important for minimizing crosstalk. This implies a need for a sufficiently large capacitance of that amplification electrode to ground. This result has consequences for how much such an amplification electrode can be segmented without incurring crosstalk.

## I. INTRODUCTION

The small electrode spacings in Micropattern Gas Detectors (MPGDs) give rise to non-negligible capacitances between various MPGD electrodes − typically on the order of tens or even hundreds of pF. Capacitive coupling among readout structures (strips or pads) as well as between readout structures and amplification structures produces signal crosstalk (XT) between adjacent and even far-away readout electrodes. In case of large signals, e.g. an MPGD avalanche produced by a heavily-ionizing particle (hip), this crosstalk can cause a large number of readout channels in the detector to go above threshold causing the entire detector to "light up" due to one localized hip event.

## II. CROSSTALK MODEL

The model created in PSpice [1] for studying the capacitive coupling of crosstalk in an MPGD is shown in Fig. 1. A voltage source V2 injects a square voltage pulse onto ganged signal strips (green). The signal strips are capacitively coupled via C5 to the facing electrode (blue), e.g. the bottom of GEM3 in Triple-GEM detectors or the mesh in Micromegas detectors. They are also coupled via the interstrip capacitance C6 to adjacent ganged strips (red) in a neighboring sector on which the crosstalk is studied. These crosstalk strips ("XT strips") also couple capacitively to the facing avalanche electrode (via C1). Signal strips, XT strips, and the facing avalanche

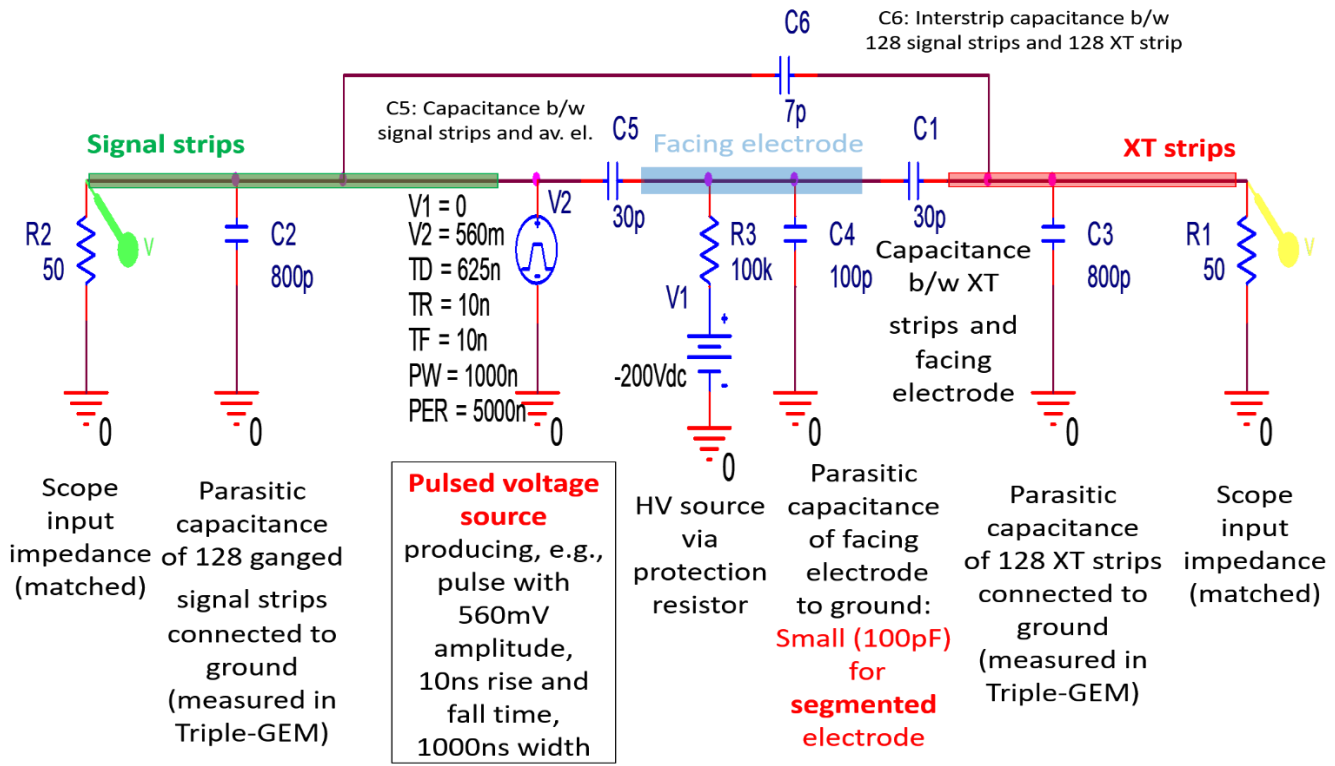

Fig. 1. PSpice circuit model for capacitive crosstalk in an MPGD.

electrode all also have parasitic capacitances (C2, C3, C4) to AC detector ground. Connections of the strips to oscilloscope channels for voltage probing are modeled by terminating them

Manuscript received December 20, 2020. This work was supported in part by the U.S. Department of Energy, Office of Science (HEP).
M. Hohlmann is with the Dept. of Aerospace, Physics and Space Sciences, Florida Institute of Technology, Melbourne, FL 32901 USA (telephone 321-674-7275, e-mail: hohlmann@fit.edu).

into 50Ω resistors (R1, R2). A DC voltage source (V1) is connected to the facing avalanche electrode via protection resistor R3 to simulate powering it with high voltage. It also acts as an AC ground for the XT current.

### III. SIMULATION RESULTS

Fig. 2 shows the simulated signal on the crosstalk strips in response to a 1μs square pulse injected onto the signal strips. The circuit acts as a CR differentiator producing clipped crosstalk pulses at rising and falling edges of the input pulse. This reproduces experimental results (Fig. 3) obtained with a Triple-GEM detector quite well [2]. The C and R values used in the simulation are selected to match the experimental configuration in [2].

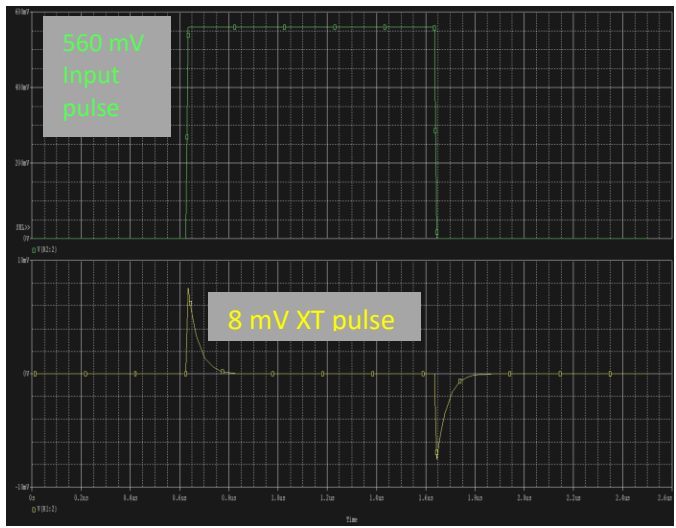

Fig. 2. Input voltage pulse (top) and output crosstalk signal (bottom) from the PSpice simulation of the circuit in Fig.1.

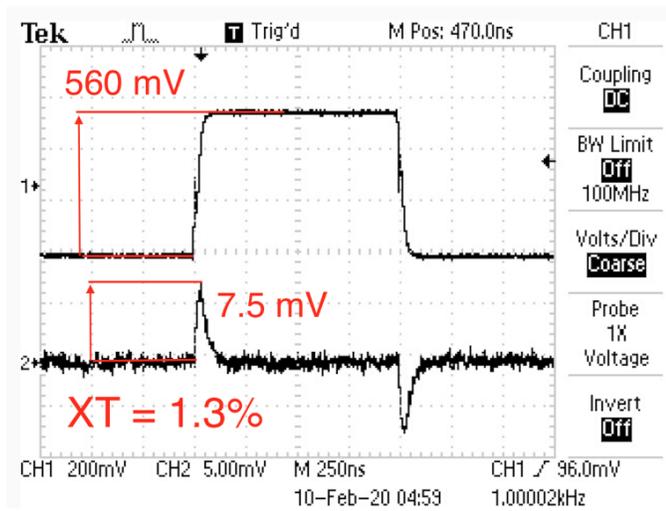

Fig. 3. Input voltage signal from a pulse generator connected to 128 ganged readout strips in a Triple-GEM detector (top) and observed crosstalk pulse on 128 ganged readout strips in a neighboring readout sector (bottom) (from [2]).

When replacing the pulsed voltage source in the simulated circuit with a pulsed *current* source to simulate actual detector operation with electron-ion avalanches, the current on the termination resistor of the signal strips has the opposite direction as the current on the termination resistor on the XT strips, i.e. signal and XT now have opposite polarities (Fig. 4), which has also been confirmed experimentally [2].

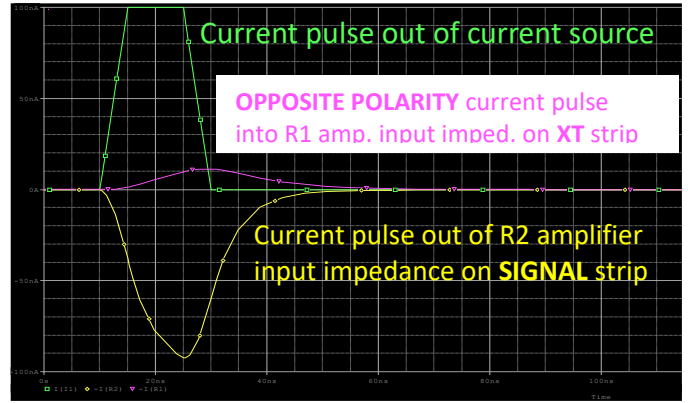

Fig. 4. Input (green), signal (yellow), and crosstalk current pulses (purple) from the PSpice simulation of the circuit in Fig.1 with a current source.

The simulation shows that the crosstalk amplitude is reduced and the shape becomes bipolar (Fig. 5) if the impedance of the facing electrode to ground is reduced, e.g. by increasing its parasitic capacitance to AC ground. In this case, more of the crosstalk AC current is sinked to ground and less reaches the XT strip. The XT pulse shape changes because it has two contributions with opposite polarity: One is the current via the capacitive coupling (C1) of the XT strip to the facing avalanche electrode, and the other is the current via capacitive coupling to the signal strips via interstrip capacitance (C6). In the idealized case of this simulation, it is actually possible to make these two contributions cancel each other out completely by appropriately choosing the C and R values in the circuit and to fully eliminate all XT current arriving at the XT strips. The observed impact of the impedance and capacitance of the facing avalanche electrode to AC ground on the XT must be carefully taken into account when segmenting this electrode.

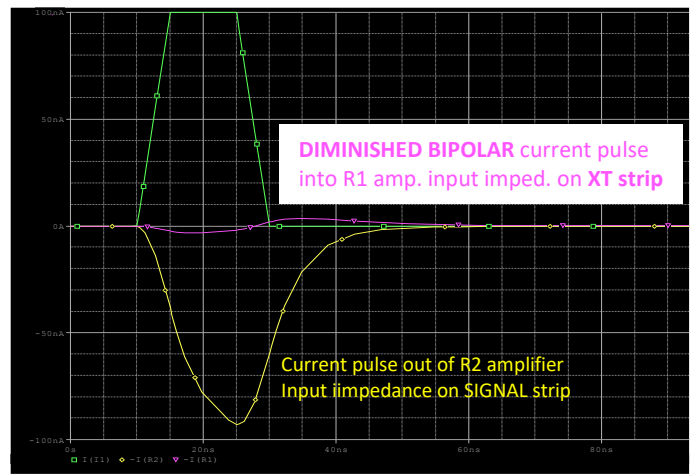

Fig. 5. Diminished crosstalk current pulse (purple) when parasitic capacitance C4 from facing avalanche electrode is increased from 100pF to 1000pF as obtained from PSpice simulation of the circuit in Fig.1 with a current source.

The simulations show that any method for reducing the impedance Z of the electrode facing the readout structure to AC GND to sink more XT current to AC GND will reduce crosstalk. For example, using a facing electrode with larger contiguous area and capacitance ($Z = 1/\omega C_4$), or connecting the facing electrode without protection resistor R3 to HV, or placing a bypass capacitor around the protection resistor R3, or placing a direct bypass capacitor from the facing electrode to GND will all achieve this objective. The simulation also shows that none of these mitigation measures have a big impact on the signal integrity on the signal strips (Fig. 5).

## IV. Summary and Conclusion

The simulation results obtained with a PSpice model of the capacitive couplings within an MPGD replicate experimental crosstalk measurements obtained for a Triple-GEM with a voltage pulser. The simulation explains why tests with a voltage pulser produce *same-sign* XT pulses while tests with a GEM in normal operation mode produce *opposite-sign* XT pulses. The reason is the different AC current flow through the capacitive network with an external voltage source compared to that with an internal current source. Reducing the impedance Z of the electrode facing the readout structure to AC GND to sink more XT current to AC GND is key to minimizing the XT in the MPGD. However, this mitigation has its limits due to the frequency dependence of the impedance and due to the inescapable presence of an additional XT path through the given interstrip capacitances.